\documentclass[preprint2]{aastex}
\usepackage{subfigure}
\usepackage{enumerate}

\usepackage{epsfig}
\usepackage{lscape}
\usepackage{color,soul}
\usepackage[normalem]{ulem}

\shorttitle{Stellar Abundancs in Binary Host System XO-2}
\shortauthors{Teske et al.}

\begin{document}

\newcommand{\txw}{\textwidth}

\title{Abundance Differences Between Exoplanet Binary Host Stars XO-2N
  and XO-2S -- Dependence on Stellar Parameters$^{*}$}


\altaffiltext{*}{Based on data collected at Subaru Telescope, which is operated by the National Astronomical Observatory of Japan.}

\author{Johanna K. Teske\altaffilmark{1,+}, Luan
    Ghezzi\altaffilmark{2}, Katia Cunha\altaffilmark{3, 4}, Verne V. Smith\altaffilmark{5}, Simon C. Schuler\altaffilmark{6}, Maria
    Bergemann\altaffilmark{7}}

\altaffiltext{1}{Carnegie DTM, 5241 Broad Branch Road, NW, Washington, DC 20015, email: jteske@carnegiescience.edu}
\altaffiltext{2}{Harvard-Smithsonian Center for Astrophysics, 60 Garden St., Cambridge, MA 02138}
\altaffiltext{3}{Steward Observatory, University of Arizona, Tucson, AZ, 85721, USA}
\altaffiltext{4}{Observat\'orio Nacional, Rua General Jos\'e Cristino, 77, 20921-400, S\~ao Crist\'ov\~ao, Rio de Janeiro, RJ, Brazil}
\altaffiltext{5}{National Optical Astronomy Observatory, 950 North Cherry Avenue, Tucson, AZ 85719, USA}
\altaffiltext{6}{University of Tampa, 401 W. Kennedy Blvd., Tampa, FL 33606, USA}
\altaffiltext{7}{Max Planck Institute for Astronomy, Koenigstuhl 17, 69117 Heidelberg, Germany}
\altaffiltext{+}{Carnegie Origins Fellow, jointly appointed by Carnegie DTM \& Carnegie Observatories}

\clearpage

\begin{abstract}

The chemical composition of exoplanet host stars is an important factor in
understanding the formation and
characteristics of their orbiting planets. The best
example of this to date is the 
planet-metallicity correlation. Other proposed
correlations are thus far
less robust, in part due to uncertainty in the chemical history of stars pre- and post-planet
formation. Binary host stars of similar type present an opportunity to isolate the effects of
planets on host star abundances. 
Here we present a
differential elemental abundance analysis of the XO-2 stellar binary,
in which both G9 stars host giant planets, one of which is
transiting. Building on our previous work,
 we report 16 elemental abundances and compare
the $\Delta$(XO-2N--XO-S) values to elemental condensation temperatures. 
The $\Delta$(N-S) values and slopes with condensation temperature resulting from four different pairs of stellar
parameters are compared to explore the effects of
changing the relative temperature and gravity of the stars. We find
that most of the abundance differences between the stars depend on the
chosen stellar parameters, but that Fe, Si, and potentially Ni are
consistently enhanced in XO-2N regardless of the chosen stellar
parameters. 
This study emphasizes the power of binary host star abundance analysis for probing the
effects of giant planet formation, but also illustrates the
potentially large uncertainties in abundance differences and slopes induced by
changes in stellar temperature and gravity.

\end{abstract}


\section{Introduction}

The first indication that stellar composition plays a role in, and may
be affected by, the
formation of planets was the enhanced metallicity (parameterized as [Fe/H]\footnote{
  [X/H]=log(N$_{\rm{X}}$/N$_{\rm{H}}$) - log (N$_{\rm{X}}$/N$_{\rm{H}}$)$_{\rm{solar}}$}) 
of solar-type exoplanet host stars (Gonzalez 1997). While initially
suggested as a sign of accretion of
hydrogen-depleted material onto the star, the planet-metallicity
correlation is now established as a mostly primordial effect (e.g., Santos et
al.\,2004; Fischer \& Valenti 2005). Recently, the correlation was
shown to be dependent on planet mass (e.g., Sousa et al. 2008; 
Ghezzi et al. 2010), being the weakest for 
terrestrial-sized planets (e.g., 
Buchhave et al. 2014). If host star metallicity is
considered a proxy for heavy element concentration in the
protoplanetary disk, this planet mass
dependence seems to support the 
core accretion process of planet formation, as
a larger concentration of heavy elements is needed to form giant
planet cores ($\sim$10 M$_E$; Ida
\& Lin 2004) rapidly enough to allow time for gaseous envelope accretion
($\gtrsim$100 M$_E$; Pollack et
al. 1996) prior to gas disk dissipation after a few million years
(e.g., Wyatt et al. 2008). 

Is it still possible that planet formation may alter host star
composition? Mel\'endez et al. (2009; M09) suggest the Sun is deficient by
$\sim$20\% in refractory elements, 
with condensation temperatures $T_c\gtrsim$ 900 K, relative to volatile
elements 
 when compared to ``solar twins'' without detected
planets. M09, as well as 
studies finding similar results 
(e.g., Ram\'irez et
al.\, 2010; Gonzalez et al. 2010), interpret the negative [X/H] or [X/Fe] vs T$_c$ slopes of
refractory elements as a star 
``missing'' rock-forming material that has been 
sequestered in terrestrial planets (Chambers
2010). However, this interpretation has been questioned. 
Other
studies have found 
no
increase of negative slopes in hotter or more metal-rich stars (Ram\'irez et al.\,2014); no difference in
slopes between hosts/non-hosts after correcting for Galactic chemical
evolution (Schuler et al. 2011b), correlations between the $T_c$ slope and stellar age
and log $g$ (Adibekyan et
al. 2014); and even the opposite of expected trends --
positive/negative slopes in small/giant planet hosts (Gonz\'alez Hern\'andez et
al.\,2011). To complicate the picture, stars hosting close-in
giant planets 
may instead accrete refractory-rich material that would have formed small planets (
e.g., Raymond et al. 2011; Schuler et al 2011b). 

Thus, if/how planet formation influences
stellar composition remains uncertain, and is challenging to
disentangle from the local or global composition of the disk and the
star's position in/motion through the Galaxy (e.g., {\"O}nehag et al. 2014; Schuler et
al. 2011b). 
Binary host stars provide an opportunity for constraining
how planet formation affects host stars, as they likely experience similar environments throughout their evolution (e.g., Kratter
2011). Desidera et al. (2004; 2006) found that the majority of $\sim$50 wide
binaries (not known to host planets) had
$\Delta$[Fe/H]$\leq$0.03, suggesting larger metallicity differences
are exceptional. Gratton et al. (2001) examined six wide binaries
across a larger range of elements, finding that four had identical
abundances. However, two stars were significantly different, with
$\Delta$[Fe/H]=0.053$\pm$0.014 and 0.091$\pm$0.006 and similar
differences for other refractory elements; the authors suggested these
differences were due to pollution by dusty protoplanetary disk or rocky
planetary material. Radial velocity (RV) monitoring of both ``polluted''
systems to search for planets has been inconclusive due to changing
stellar activity cycles. Within binary systems known to host planets,
the results are also ambiguous: There is not yet consensus on the significance of
differences between 16 Cyg A and B (the latter hosts a $\sim$2.4 M$_J$
planet; Schuler et al. 2011a; Ram\'irez et al.\,2011; Tucci Maia
et al. 2014), and similar studies indicate no abundance
differences between stars in a single-host (HAT-P-1, Liu et al. 2014) and dual-host (HD
20782/1, Mack et al. 2014) systems. 
 
Recently, RV monitoring of the southern component (XO-2S; 0.98$\pm$0.05
M$_{\odot}$; Desidera et al. 2014) of
the $\sim$4600 AU binary XO-2 revealed two giant (0.26 M$_J$, 1.37
M$_J$), slightly eccentric ($e \sim$0.18, 0.15) exoplanets at 0.13 and
0.48 AU (Desidera et al. 2014). XO-2N (0.98$\pm$0.02 M$_{\odot}$;
Burke et al. 2007) was already known to host a
transiting hot Jupiter (0.62 M$_J$, $e$=0, $a$= 0.04 AU). This system is one
of only four known dual-planet-hosting binaries (XO-2, HD 20782/1, Kepler-132, and WASP-94). 
Here we expand the analysis of Teske et
al. (2013) 
to investigate whether there are 
abundance differences between the stars, how such differences depend
on derived stellar parameters, and how they may relate to the formation of different types of
planets. 

\section{Observations and Data Analysis}
The observations and data reduction for XO-2N and -2S are detailed in Teske et
al. (2013). The $R \sim$60,000, S/N$\sim$170-230 data were acquired in February 2012 with the High
Dispersion Spectrograph 
at the 8.2 m Subaru
Telescope, and reduced using standard techniques with the
IRAF\footnote{IRAF is distributed by the National Optical Astronomy
  Observatory, which is operated by the Association of Universities
  for Research in Astronomy, Inc., under cooperative agreement with
  the National Science Foundation.} software package. 

\subsection{Original Stellar Parameters}
The stellar parameter analysis and uncertainty calculations are also explained in Teske et
al. (2013) and listed in Table \ref{tab:stellar_params}. 
The abundances were determined using the local thermodynamic equilibrium (LTE) spectral analysis code
MOOG (Sneden 1973), with model atmospheres interpolated from the
Kurucz ATLAS9 NOVER grids\footnote{See
  http://kurucz.harvard.edu/grids.html.}. Equivalent widths were
measured with `splot' in IRAF, 
and abundances were 
normalized to solar
, derived from a reflected-moonlight
spectrum taken with the same configuration during the same run, on a
line-by-line basis. In XO-2N, 49 Fe I and 8 Fe II lines were measured;
in XO-2S, 49 Fe I and 10 Fe II lines were measured; the line
properties and equivalent widths are
provided in Table \ref{tab:lines}. 



\subsection{Alternative Stellar Parameters}
As shown in recent studies of GAIA benchmark stars (Jofre et al. 2014)
and transiting exoplanet host stars (e.g., Torres et al. 2012), different
analysis methods lead to different stellar parameters and thus
different abundances. 
To investigate this effect thoroughly, 
particularly the different T$_{\rm{eff}}$s, we explored several 
alternative methods, used widely in other exoplanet host star studies, for deriving stellar parameters for XO-2N and -2S.

\textbf{Alternative Parameter Set 1} First, we repeated our analysis relative to solar, using by-hand IRAF
measurements of the lines in Ghezzi et al. (2010). This list includes a
smaller number of Fe I and II lines (in our case 23 Fe I for N/S,
and 9/10 Fe II for N/S), but all have
lab-measured and thus more reliable gf-values. Kurucz ATLAS9 ODFNEW
grids were used as input models, though the difference between NOVER
and ODFNEW are insignificant. The resulting
parameters are listed as ``Alt Param 1'' in Table
\ref{tab:stellar_params}. 

\textbf{Alternative Parameter Set 2} Second, we adopted the full analysis approach of Ghezzi et al. (2010) and
derived parameters based on absolute abundances, not relative
to solar. The Ghezzi et al. (2010) line list, with lab-measured
gf-values, was compiled specifically for an absolute
analysis. Again, the measurements were done by-hand in IRAF and the
stellar models were ODFNEW, using the same EWs as ``Alt Param 1''. The resulting
parameters, are listed as ``Alt Param 2'' in Table
\ref{tab:stellar_params}. 

\textbf{Alternative Parameter Set 3} Third, we used ODFNEW models with a different and longer line
list of 72/67 Fe I and 9/8 Fe II lines for N/S, compiled from Sousa et
al. (2008). The equivalent widths were measured with ARES (Sousa et
al. 2007); lines $>$0.30 dex from the average in the Sun were removed
and an interative 2$\sigma$ clipping was applied to the stars during the parameter determination. The resulting parameters are listed as ``Alt Param 3'' in Table
\ref{tab:stellar_params}. 

\subsection{Parameter Comparison}
In almost every case, within 1$\sigma$ total errors all derived
parameter pairs for XO-2N and
-2S overlap, indicating the stars are
similar, potentially identical, but could also differ by as much as $\sim$200
K in T$_{\rm{eff}}$, $\sim$0.25 dex in log $g$, $\sim$0.10 in [Fe/H],
and $\sim$0.07 km/s in $\xi$. In the first three parameter sets,
T$_{\rm{eff}}$ of XO-2N $<$ XO-2S, which agrees with the original
XO-2Nb discovery paper (Burke et al. 2007), T$_{\rm{eff}}$s calculated from the IRFM photometric calibration of Casagrande et al. (2010) using colors
from Burke et al. (2007), and the recent analysis of HARPS-N spectra by Damasso et al. (2015). The original and alternative parameter set 3
both find log $g_{\rm{N}} >$ log $g_{\rm{S}}$, while
alternative parameters 1 and 2 find the opposite; however, given the log
$g$ errors, the magnitude of $\Delta$log $g$ between the two
stars is uncertain (and could be zero). Interestingly, a higher [Fe/H]
in XO-2N is consistent across all derived parameters; this also
concurs with Damasso et al. (2015). Despite the possibility of
the stars having the same [Fe/H] within errors, the fact that all four
parameter sets -- found using different lines measurements and analyses techniques -- find [Fe/H]$_{N}
>$[Fe/H]$_{S}$ points towards a real metallicity difference. We return
to this point in \S 3. 


\subsection{Other Elemental Abundances}

All abundances 
were derived directly from equivalent
width (EW) measurements performed with 
the `splot' task in IRAF, except [O/H], which was derived through spectral
synthesis in MOOG. Abundance
determinations of Fe, Ni, C, and O are detailed in Teske et
al. (2013).
For elements new to this work
, line lists (Table \ref{tab:lines}) were compiled from Mel\'endez et
al. (2014) and Schuler et al. (2011b); atomic parameters are
from numerous sources and compiled in VALD (Piskunov et
al. 1995). All $\Delta$(N-S) abundance differences are listed in Table \ref{tab:diffs}.


The only element besides Fe for which we measured two ionization
states was titanium. 
We removed 1$\sigma$ outliers
based on the mean and standard deviation of the line-by-line
solar-normalized abundances. 
This procedure changed the [Ti I/H] and [Ti II/H] values by
$\leq$0.025, but reduced the dispersion in the abundances by
roughly a factor of two. 

Bergemann (2011) found the discrepancy between solar Ti I
and II is solved by including non-LTE line
formation, and suggested that Ti II may be a better [Ti/H] indicator
in solar-type stars. 
We tested whether non-LTE corrections to Ti I
lines in XO-2N and -2S, calculated using
 statistical equilibrium and line formation codes described in
 Bergemann (2011) and MAFAGS-OS models with the same stellar
 parameters derived here, change the [Ti I/H] values. We find these corrections do not
 significantly alter [Ti I/H], but as these are the first
 non-LTE corrections calculated for cool, metal-rich stars, we provide the
 A(Ti I)$_{NLTE}$-A(Ti I)$_{LTE}$ corrections in Table
 \ref{tab:lines}. We 
report a total [Ti/H]
based on the lines of both species, with 1$\sigma$ outliers
removed within each species as described above. 

\section{Results and Discussion}

The results of our measurements are in
Table \ref{tab:lines}, including wavelength, $\chi$, log $gf$,
EWs for each element for the Sun, XO-2N,
and XO-2S. 
The final derived stellar parameters and
their 1$\sigma$ uncertainties are in
Table \ref{tab:stellar_params}, and the
$\langle\Delta$(X/H)$\rangle$ and $\sigma$ values calculated on a
line-by-line basis for XO-2S--XO-2S are in Table \ref{tab:diffs}. 
Next we discuss the resulting abundances from the four
different parameter sets derived above. 

\subsection{Relative Abundances versus T$_c$ }

\begin{figure}[ht!]
\centering
\subfigure{\includegraphics[width=0.75\textwidth]{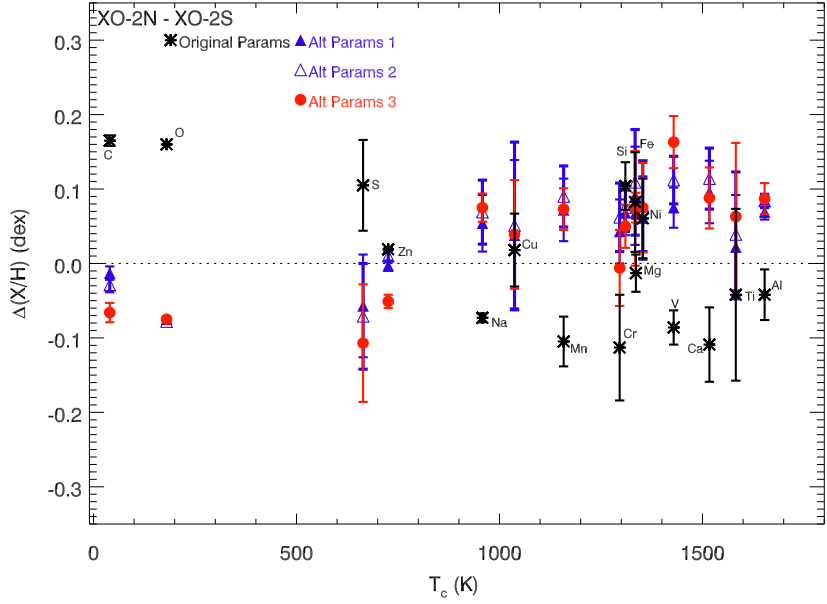}}
\quad
\subfigure{\includegraphics[width=0.75\textwidth]{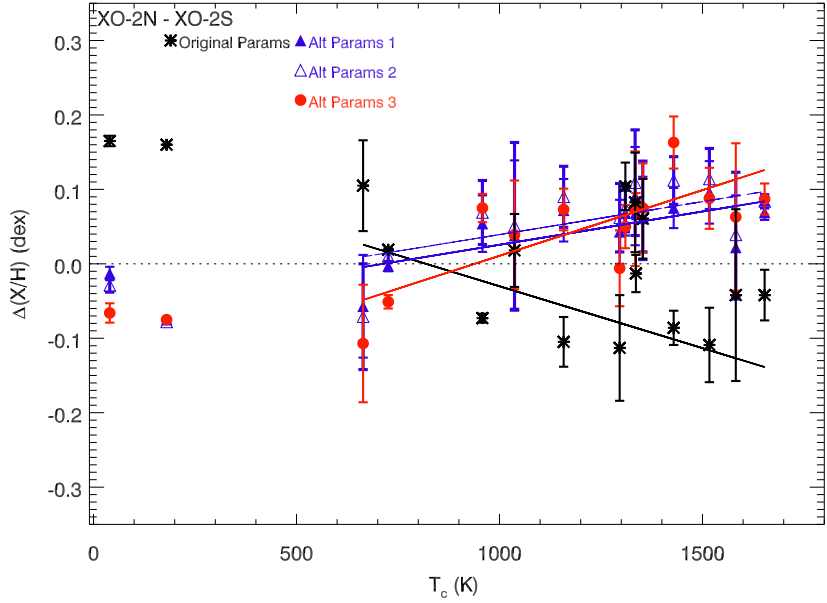}}
\caption{\textbf{Top:} The $\Delta$(XO-2N - XO-2S) relative
  abundances versus T$_c$ (Lodders 2003). Black asterisks, blue closed/open
  triangles, and red circles show results from using the ``original'' parameters,
  ``alternative params 1'', ``alternative params 2'', and
  ``alternative params 3'', respectively. A
  dotted line shows zero difference. All models show a Si, Fe, and possibly Ni enhancement in XO-2N. \textbf{Bottom:}
  Same as top, showing fits to T$_c$$\ge$500 K elements; the
  dashed blue line corresponds to the open triangle points. The red/blue fits have positive slopes, whereas the black fit
has a negative slope.}
\label{fig2}
\end{figure}

The advantage of studying binary stars is the likelihood that
their formation environments were similar, so
their compositions -- assuming like XO-2N and -2S they are of similar mass and temperature -- should also be similar. Small differences in binary star
abundances were previously suggested to be related to planet
formation/migration (e.g., Tucci-Maia et al. 2014). To investigate the
XO-2 system for such abundance differences, we compare one star
directly to the other, $\Delta$(XO-2N - XO-2S).
In Figure \ref{fig2} we show the line-by-line (N-S)
differences 
and standard
deviations of the differences (error bars) for our four parameter sets.

The black asterisks in Figure \ref{fig2} represent the 
largest possible differences between (N-S) --
-204 K $\Delta$T$_{\rm{eff}}$, 0.27 dex $\Delta$log $g$, and 0.11 dex
[Fe/H]. Based on these differences, XO-2N may have
less refractories than XO-2S, save a few
elements -- Si, Fe, and Ni. These ``original'' parameter results also
imply XO-2N is enhanced in volatile elements; the slope for T$_c
\geq$500 K elements is negative ($-$1.66E-4 dex/K).

However, the abundance differences, and thus slopes, change with different stellar
parameters, as suggested in \S 2. Using a smaller line list, relative-to-solar/absolute
results in the blue closed/open triangles in Figure \ref{fig2}. These
parameters correspond to (N-S) differences of -83/-72 K $\Delta$T$_{\rm{eff}}$,
-0.09/-0.11 dex $\Delta$log $g$, and 0.09/0.11 dex
[Fe/H]. Interestingly the relative-to-solar and absolute analyses
produce a very similar pattern, indicating that XO-2N is enhanced in most refractory elements and
depleted in volatiles. The red points in Figure \ref{fig2} represent abundances represent parameters derived
from a longer line list than the ``original'' parameter set, with
measurements made with ARES, and (N-S)
differences of 3 K $\Delta$T$_{\rm{eff}}$, 0.06 dex $\Delta$log $g$, and 0.08 dex
[Fe/H]. These abundance differences show an even greater
enhancement/depletion in XO-2N of refractory/volatile elements versus
the abundances marked with blue triangles. 

The different $\Delta$(N-S) abundances, and corresponding T$_c$-slopes, resulting from different
stellar parameters make it challenging to conclude whether any of the
enhancements/depletions between XO-2N and -2S are real, within the
bounds of measurement errors. 
To investigate how dependent the abundance differences are on
$\Delta$T$_{\rm{eff}}$ and $\Delta$log $g$, which change the most between our
different analyses, we used the same EWs and
three test parameter sets that varied only in $\Delta$log $g$. For
these tests, we adopted $\Delta$T$_{\rm{eff}}$ -130 K (N-S), based
on the spread of the points in Figure \ref{fig2}; this
$\Delta$T$_{\rm{eff}}$ is between the $\Delta$T$_{\rm{eff}}$s of the
black asterisks (-204 K) and red circles (3 K). Specifically, the
parameters for XO-2N/S were 5390/5520 K, 0.43/0.33 dex [Fe/H], and
1.20 km/s for $\xi$, with log $g$ varying between 4.30/4.40, 4.35/4.35,
and 4.40/4.30.  

\begin{figure}[ht!]
\centering
\includegraphics[width=1.0\textwidth]{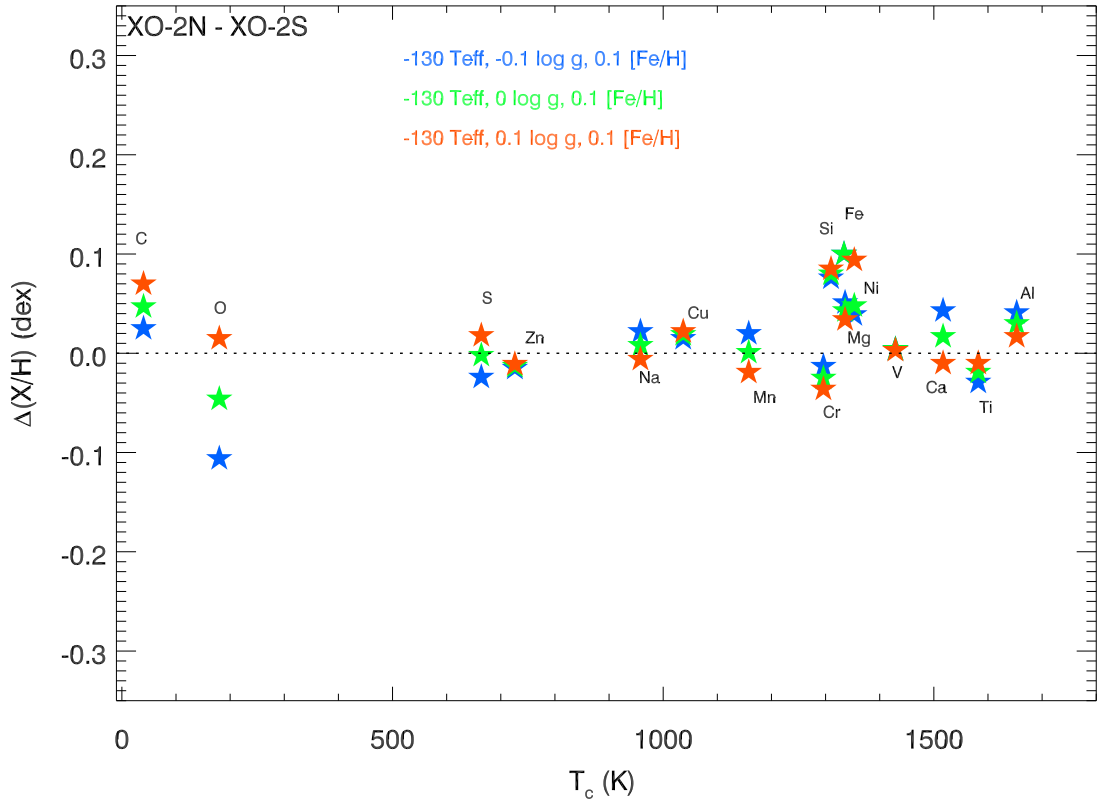}
\caption{The $\Delta$(XO-2N - XO-2S) relative
  abundances versus T$_c$ for our test cases, keeping
  $\Delta$T$_{\rm{eff}}$ for (N-S) at -130 K, $\Delta$[Fe/H] for (N-S)
  at 0.1 dex, and varying $\Delta$ log $g$ (N-S) from -0.1 (blue stars), to 0
  (green stars), to +0.1 (orange stars) dex. A dotted line shows zero difference. }
\label{fig3}
\end{figure}

The results of changing $\Delta$log $g$ between XO-2N and -2S are
shown in Figure \ref{fig3} as blue (-0.1 dex), green (0 dex), and
orange (0.1 dex) stars.Changing $\Delta$log $g$
has a minimal effect on the abundance differences, except for
$\Delta$(O/H), which is known to depend strongly on log $g$. For most
elements, the log $g$ tests result in $\Delta$(N-S) abundances of $\sim$0, falling somewhere in between the black, blue, and
red points from Figure \ref{fig2}. 

However, in all three tests in Figure \ref{fig3}, the abundances of
Si, Fe, and Ni are still above the zero difference line. This is remarkably consistent with all four
parameter sets derived in \S 2 and plotted in Figure \ref{fig2},
which all show Si, Fe, and possibly Ni enhancements in XO-2N. Varying many
different aspects of the analysis and stellar parameters does not
alter this enhancement pattern, providing strong evidence for a real
abundance difference in Si, Fe, and Ni between XO-2N and -2S.

\subsubsection{Possible Causes of Abundance Pattern}

Robinson et al. (2006) and Brugamyer et al. (2011) found differences
in [Si/Fe] between known host and non-host stars; 
 Robinson et al. (2006) also found that host stars are enhanced in
nickel. While these studies compare known hosts and non-hosts,
they point specifically at Si and Ni as important to giant planet
formation. This is interesting given these are two elements we
find enhanced in XO-2N, the hot Jupiter host. Indeed, plotting Brugamyer et
al. (2011)'s [Si/H] values against planet semi-major axis shows a
dearth of silicon-poor hosts to close-in ($a<0.1$ AU) planets, suggesting that silicon may play a role in
close-in giant planet formation.


The enrichment of certain elements in XO-2N could be due to accretion of material preferentially enhanced in
these elements. Simulations have shown
that a significant fraction of giant planets migrating in disks as a
result of planet-planet scattering cause much, if not all, of 
rocky disk material to be accreted onto the star (Raymond et
al. 2011). The RV evidence for a second object orbiting XO-2N (Damasso et
al. 2015; Knutson et al. 2014; Narita et al. 2011) adds weight to the planet-planet scattering
migration scenario (Rasio \& Ford 1996; Nagasawa et al. 2008),
 by which XO-2Nb could have
migrated inward and subsequently had its eccentricity and spin-orbit
alignment angle damped (Narita et al. 2011).


XO-2S's lesser amount of Fe, Si, and Ni may instead indicate
that these elements are important to \textit{small} planet
formation; XO-2S may have 
formed small planets 
that were ejected. 
 Kaib et al. (2013) specifically studied wide
binaries, finding that $\sim$70\% of 1 M$_{\odot}$-1 M$_{\odot}$ systems with
the separation of XO-2N and -2S were unstable and ejected a planet
within 10 Gyr. 
Based on dynamical simulations of wide binary planet systems, Morbidelli (2014) found that 73\% of
planets are ejected, and that 60\% of wide binaries, versus 41\% of isolated systems, lost a planet. 

\subsection{Comparison to Other Host Star Binary Studies}
In all parameter cases, a few refractory elements (Si, Fe, Ni) stand
out as enhanced in the cooler XO-2N. 
This pattern is different from other abundance studies of binary exoplanet
hosts. Liu et al. (2014) found $\Delta$[X/H] values consistent with
zero in the HAT-P-1 G0+F8 binary. Though there is not consensus
regarding 
differences between 16 Cyg A (G1.5) and
B (G3), the proposed differences (Ram{\'{\i}}rez et al. 2011; Tucci Maia et
al. 2014) either show $\sim$0.04 dex uniform depletion with T$_c$, or a range of depletions
increasing with T$_c$ at the -0.03 to -0.05 dex level. (We switched A-B values to B-A, to stay consistent
with cooler-hotter star.) Mack et al. (2014), who measured the
double-host binary HD20781/HD20782 (G9.5/G1.5), show in their Figure 3 non-zero
differences for O, Al, V, Cr, Fe, and Co.  Are we seeing an effect in $\Delta$[X/H] trends due to
T$_{\rm{eff}}$, moving from hotter stars and no
differences to cooler stars and more significant differences? Right now this question is hard to
answer due to different analysis techniques, line lists, and
the host star nature of these systems (whether or not both stars host
planets). In Gratton et al. (2001)'s Figure 1, plotting
$\Delta$[Fe/H] versus $\Delta$T$_{\rm{eff}}$ for stars without
confirmed planets,  there is no clear
pattern -- $\Delta$[Fe/H] is near zero for
stars with $\Delta$T$_{\rm{eff}}$ $\sim$10, 90, 275, and 330 K, and $\Delta$[Fe/H] = 0.05 and 0.09 for stars with
$\Delta$T$_{\rm{eff}}$ $\sim$120 and 275 K. Desidera et al. (2006)'s
data hint at larger $\Delta$[Fe/H] differences at larger
$\Delta$T$_{\rm{eff}}$s, but Desidera et al. (2004)'s data show no
real dependence of $\Delta$[Fe/H] on $\Delta$T$_{\rm{eff}}$. More
uniform abundances studies, and a larger sample,
are needed to understand the temperature dependence of binary
host star abundance differences and whether they are related to
planet formation.


\section{Conclusions}

We performed a detailed stellar abundance analysis on XO-2N and XO-2S, binary stars of similar temperature and mass that host planets of different
types, to look for effects of planet formation. Though binary
stars are expected to have approximately identical abundances, some studies report differences between host
stars that may be due to planet formation. 


When comparing the stars relative to one
another, 
there are hints of abundance
differences. We test how dependent these differences, and the
$\Delta$(X/H) slopes
with T$_c$, are on stellar
parameters using two different line lists, solar-normalized
and absolute abundance analyses, and measurements made by hand in IRAF
and automatically with ARES. Most of the resulting $\Delta$(X/H) values vary
widely, but a few refractory elements -- Si, Fe, and possibly Ni -- consistently
appear to be enhanced in XO-2N. This pattern is independent of
analysis method and adopted T$_{\rm{eff}}$ and log $g$ value, giving
confidence to the conclusion that XO-2N and -2S have real abundance differences.

This work adds to the few binary host star analyses, which may prove to be the most useful for
constraining both the effects of planet formation on stellar
abundances, and the bulk compositions of giant exoplanets. However,
it also illustrates how sensitive binary host star
abundances and T$_c$ slopes are to derived stellar parameters; such studies approach
the limit of measurement errors and should be considered carefully. 


\acknowledgements
The authors wish to recognize and acknowledge the very significant
cultural role and reverence that the summit of Mauna Kea has always
had within the indigenous Hawaiian community. We are most fortunate to
have the opportunity to conduct observations from this mountain. 
We thank the referee for their comments and edits that improved the
paper. 
{\it Facilities:} \facility{Subaru}


\begin{landscape}
\begin{deluxetable}{lccccccccccc}
\tabletypesize{\scriptsize}
\tablecolumns{12}
\tablewidth{0pc}
\tablecaption{Derived Stellar Parameters and $\Delta$(N-S) Abundances \label{tab:stellar_params}}
\tablehead{ 
\colhead{Parameters} & \colhead{ T$_{\rm{eff}}$ (K)} & \colhead{error} & \colhead{ log $g$ (cgs) }  & \colhead{error}&
\colhead{$\xi$ (km s$^{-1}$)}  & \colhead{error} &
\colhead{$\rm{[Fe/H]}$} & \colhead{error} & \colhead{line list} &
\colhead{relative to solar?} & \colhead{IRAF/ARES}}
\startdata
Orig Params, XO-2S & 5547 & 59 & 4.22 & 0.24 & 1.24 & 0.07 & 0.28 &0.14 & 49 Fe II, 10 Fe II & yes & IRAF  \\
Orig Params, XO-2N & 5343 & 78 & 4.49 & 0.25 & 1.22 & 0.09 & 0.39 &
0.14 & 49 Fe II, 8 Fe II & yes & IRAF \\
\hline
Alt Params 1, XO-2S & 5523 & 49 & 4.44 & 0.11 & 1.35 & 0.08 & 0.36 &
0.04 & 23 Fe I, 10 Fe II & yes & IRAF \\
Alt Params 1, XO-2N & 5440 & 69 & 4.35 & 0.19 & 1.29 & 0.09 & 0.45 &
0.06 & 23 Fe I, 9 Fe II & yes & IRAF \\
\hline
Alt Params 2, XO-2S & 5447 & 47 & 4.36 & 0.11 & 1.08 & 0.06 & 0.28 &
0.04 & 23 Fe I, 10 Fe II & no & IRAF \\
Alt Params 2, XO-2N & 5375 & 78 & 4.25 & 0.20 & 1.02 & 0.11 & 0.39 &
0.07 & 23 Fe I, 9 Fe II & no & IRAF \\
\hline
Alt Params 3, XO-2S &5403 & 39 & 4.40 & 0.13 & 1.06 & 0.07 & 0.39 &
0.06 & 67 Fe I, 8 Fe II & yes & ARES \\
Alt Params 3, XO-2N & 5406 & 32 & 4.46 & 0.08 & 1.11 & 0.05 & 0.47 &
0.05 & 72 Fe I, 9 Fe II & yes & ARES \\
\enddata
\end{deluxetable}
\end{landscape}

\begin{deluxetable}{lcccccccccc}
\tablecolumns{7}
\tablewidth{0pc}
\tablecaption{Measured Lines \& Equivalent Widths \label{tab:lines}}
\tablehead{ \colhead{Ion} & \colhead{$\lambda$} & \colhead{$\chi$} & \colhead{log $gf$} & \colhead{EW$_{\odot}$} & \colhead{XO-2S} &\colhead{XO-2N} \\ 
  \colhead{ } & \colhead{({\AA})} & \colhead{(eV)} & \colhead{(dex)} & \colhead{(m{\AA})} & \colhead{EW (m{\AA})} & \colhead{EW ({\AA})}}
\startdata
C I & 5052.17 & 7.86 & -1.304 & 33.9 & 36.4 & 33.0 \\ 
{ } & 5380.34 & 7.69 & -1.615 & 19.4 & 22.4  & 20.5  \\ 
O I & 6300.30 & 0.00 & -9.717 & \nodata & \nodata & \nodata  \\
Na I & 4751.82 & 2.10 & -2.078 & 13.1 &  \nodata & 48.0  \\ 
{} &    5148.84 & 2.10 & -2.044 & 12.4&  42.0 & 45.8 \\
{} &    6154.23 & 2.10 & -1.547 & 36.2 & 82.9 & 89.3  \\
{} &   6160.75 & 2.10 & -1.246 & 55.4  & 103.7 & 113.0  \\
\enddata
\tablecomments{This table is available in its entirety in a machine-readable form online.} 
\end{deluxetable}

\begin{deluxetable}{lcccccccc}
\tabletypesize{\scriptsize}
\tablecolumns{9}
\tablewidth{0pc}
\tablecaption{Derived $\Delta$(N-S) Abundances \label{tab:diffs}}
\tablehead{ 
\colhead{$\Delta$(X/H)} & \colhead{ Orig $\langle$N-S$\rangle$} & \colhead{ Orig $\sigma$(N-S)} &
\colhead{ AP 1 $\langle$N-S$\rangle$} & \colhead{ AP 1 $\sigma$(N-S)}  &
\colhead{ AP 2 $\langle$N-S$\rangle$} & \colhead{ AP 2 $\sigma$(N-S)} &
\colhead{ AP 3 $\langle$N-S$\rangle$} & \colhead{ AP 3 $\sigma$(N-S)}}
\startdata
$\Delta$(C/H) & 0.165 & 0.007 & -0.013 & 0.009 & -0.029 & 0.009 & -0.066& 0.013  \\
$\Delta\rm{(O/H)}$ & 0.16 & \nodata & -0.075 & \nodata & -0.078 &\nodata &  -0.075 & \nodata\\
$\Delta\rm{(Na/H)}$ & -0.073 &  0.006 & 0.054 & 0.038 & 0.069 & 0.043 & 0.075&0.019 \\
$\Delta\rm{(Mg/H)}$ & -0.013 &  0.025 & 0.066 & 0.009 & 0.079 & 0.008 &  0.075 & 0.020  \\
$\Delta\rm{(Al/H)}$ & -0.042 &  0.066 & 0.069 & 0.010 & 0.084 & 0.009 &  0.087 & 0.021  \\
$\Delta\rm{(Si/H)}$ &0.104&  0.032 & 0.068 & 0.030 & 0.075 & 0.032 &    0.050 & 0.029 \\
$\Delta\rm{(S/H)}$ & 0.105 &  0.061 & -0.057 & 0.069 & -0.071 & 0.071 &  -0.107 &  0.079\\
$\Delta\rm{(Ca/H)}$ & -0.109 &  0.050 & 0.096 & 0.042 & 0.114 & 0.041 & 0.088 & 0.041 \\
$\Delta\rm{(Ti/H)}^{a}$ & -0.042 &  0.115 & 0.022 & 0.070 & 0.039 & 0.084 &  0.063 & 0.099 \\
$\Delta\rm{(V/H)}$ & -0.086 &  0.023 & 0.075 & 0.027 & 0.112 & 0.032 &    0.163  & 0.035 \\
$\Delta\rm{(Cr/H)}$ & -0.113 &  0.071 & 0.043 & 0.043 & 0.062 & 0.046&  -0.006 &0.051  \\
$\Delta\rm{(Mn/H)}$ & -0.105 & 0.033 & 0.072 & 0.042 & 0.090 & 0.041 & 0.073 & 0.028 \\
$\Delta\rm{(Fe/H)}^{b}$ & 0.083 & 0.067 & 0.091 & 0.066 & 0.109 & 0.071 &   0.074& 0.077  \\
$\Delta\rm{(Ni/H)}$ & 0.061 & 0.054 & 0.061 & 0.056 & 0.077 & 0.061&  0.075& 0.060 \\
$\Delta\rm{(Cu/H)}$ & 0.018 & 0.049 & 0.038 & 0.101 & 0.051 & 0.112 &  0.039& 0.073  \\
$\Delta\rm{(Zn/H)}$ & 0.019 & 0.002 & -0.003 & 0.007 & 0.010 & 0.008  &  -0.051& 0.009  \\
\enddata

\tablenotetext{a}{Combined  Ti I and II average.} 
\tablenotetext{b}{Combined Fe I and II average.} 
\end{deluxetable}



\onecolumn


\newpage

\end{document}